\documentclass[12pt, draftclsnofoot, onecolumn]{IEEEtran}
\IEEEoverridecommandlockouts
\usepackage{lineno}
\usepackage{cite}
\usepackage{hyperref}
\usepackage{amsmath,amssymb,amsfonts}
\usepackage{amsmath}
\usepackage{amsthm}

\usepackage{graphicx}
\usepackage{textcomp}
\usepackage{xcolor}
\usepackage{graphicx}
\usepackage{float}
\usepackage{subfigure}
\usepackage{amsmath}
\usepackage{amsfonts,amssymb}
\usepackage{mathrsfs}
\usepackage{mathtools}
\usepackage{algorithm}
\usepackage{algorithmicx}
\usepackage{algpseudocode}
\usepackage{bm}
\usepackage{multirow}
\usepackage{array}
\usepackage{amssymb}
\usepackage{amsmath}
\usepackage{cite}
\usepackage{url}
\usepackage{xcolor}
\usepackage{cite,graphicx,amsmath,amssymb}
\usepackage{subfigure}
\usepackage{fancyhdr}
\usepackage{mdwmath}
\usepackage{mdwtab}
\usepackage{caption}
\usepackage{amsthm}
\usepackage{setspace}
\usepackage{bm}
\usepackage{algorithm}
\usepackage{algpseudocode}
\usepackage{mathtools}
\usepackage{dsfont}
\usepackage{bbm}
\newtheorem{remark}{Remark}
\newtheorem{theorem}{Theorem}

\newtheorem{lemma}{Lemma}

\newtheorem{corollary}{Corollary}

\makeatletter
\newcommand{\biggg}{\bBigg@{3}}
\newcommand{\Biggg}{\bBigg@{3.5}}
\makeatother
\begin{document}
\title{MIMO-ISAC: Performance Analysis and Rate Region Characterization}
\author{Chongjun~Ouyang, Yuanwei~Liu, and Hongwen~Yang
\thanks{C. Ouyang and H. Yang are with the School of Information and Communication Engineering, Beijing University of Posts and Telecommunications, Beijing, 100876, China (e-mail: \{DragonAim,yanghong\}@bupt.edu.cn).}
\thanks{Y. Liu is with the School of Electronic Engineering and Computer Science, Queen Mary University of London, London, E1 4NS, U.K. (e-mail: yuanwei.liu@qmul.ac.uk).}
}
\maketitle

\begin{abstract}
This article analyzes the performance of sensing and communications (S\&C) achieved by a multiple-input multiple-output downlink integrated S\&C (ISAC) system. Three ISAC scenarios are analyzed, including the sensing-centric design, communications-centric design, and Pareto optimal design. For each scenario, diversity orders and high signal-to-noise ratio slopes of the sensing rate (SR) and communication rate (CR) are derived to gain further insights. Numerical results reveal that \romannumeral1) ISAC achieves the same diversity order as existing frequency-division S\&C (FDSAC) techniques; \romannumeral2) ISAC achieves larger high-SNR slopes and a broader SR-CR region than FDSAC.
\end{abstract}

\begin{IEEEkeywords}
Integrated sensing and communications (ISAC), performance analysis, sensing-communication rate region.	
\end{IEEEkeywords}

\section{Introduction}
Integrated sensing and communications (ISAC) is a promising enabler for the development of the six-generation (6G) and beyond wireless networks \cite{Liu2022_JSAC}. The main feature of ISAC is to allow communications and sensing to share the same time-frequency-power-hardware resources. Compared with existing frequency-division sensing and communications (FDSAC) techniques, in which sensing and communications (S\&C) require isolated frequency bands as well as hardware infrastructures, ISAC is envisioned to be more spectrum-, energy-, and hardware-efficient \cite{Liu2022_JSAC,Zhang2021_JSTSP,Wang2022_CL}. Due to these attractive characteristics, ISAC has received considerable attention from both industry and the research community \cite{Liu2022_JSAC,Zhang2021_JSTSP,Yuan2021_TVT,Ouyang2022_WCL,Tang2019_TSP,Liu2022_TSP,Wang2022_CL}.

From the perspective of information, the S\&C performance of ISAC systems is evaluated by two metrics including the sensing rate (SR) and the communication rate (CR) \cite{Zhang2021_JSTSP}. Specifically, the SR measures how much environmental information can be extracted from the sensing echoes, whereas the CR measures how much data information can be recovered from the received symbols \cite{Zhang2021_JSTSP,Yuan2021_TVT,Ouyang2022_WCL}. These two metrics evaluate the fundamental information-theoretic limits of the S\&C performance achieved by ISAC. With this in mind, the authors in \cite{Yuan2021_TVT} optimized the weighted sum of the CR and the SR in a downlink multiple-input multiple-output (MIMO) ISAC system with a single communication user terminal (UT). This work focused more on the dual-functional S\&C (DFSAC) precoding design and neglected the discussion of basic system insights. As an advance, the authors in \cite{Ouyang2022_WCL} extended the work in \cite{Yuan2021_TVT} to a multiuser case and discussed the high signal-to-noise ratio (SNR) slopes of the CR and SR as well as an inner bound of the SR-CR region. Although these two works have made great progress in understanding the SR and CR in ISAC systems, there are still many important unsolved problems. For example, the Pareto boundary, i.e., the upper-right boundary of the rate region that contains all the achievable SR-CR tuples \cite{Zhang2010_TSP}, has not been characterized. For another example, a rigorous comparison between the SR-CR regions achieved by ISAC and FDSAC is still missing. It is worth noting that the Pareto boundary and rate region can be exploited to evaluate the S\&C performance limit. Hence, characterizing the Pareto boundary and rate region is of great theoretical importance in understanding the superiority of ISAC over FDSAC.

This letter analyzes the S\&C performance of a downlink MIMO-ISAC system. The main contributions are summarized as follows: \romannumeral1) We provide DFSAC precoding design for three scenarios, including the sensing-centric (S-C) design (maximizing the SR only), the communications-centric (C-C) design (maximizing the CR only), and the Pareto optimal design (characterizing the Pareto boundary of the SR-CR region); \romannumeral2) For each scenario, we analyze the outage probability (OP) of the CR and show that ISAC yields the same diversity order as FDSAC; \romannumeral3) For each scenario, we derive the high-SNR slopes of the CR and SR and show that ISAC provides larger high-SNR slopes and thus more degrees of freedom than FDSAC in terms of both sensing and communications; \romannumeral4) We provide a rigorous comparison between the SR-CR regions achieved by ISAC and FDSAC and prove that ISAC achieves a broader SR-CR region than FDSAC.

\begin{figure}[!t]
\centering
\setlength{\abovecaptionskip}{0pt}
\includegraphics[height=0.3\textwidth]{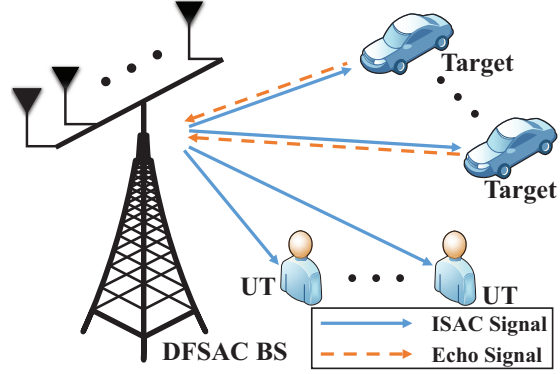}
\caption{Illustration of a downlink MIMO-ISAC system.}
\label{Figure1}
\vspace{-20pt}
\end{figure}
\vspace{-10pt}
\section{System Model}
\vspace{-5pt}
Consider a downlink MIMO-ISAC system as shown in {\figurename} {\ref{Figure1}, where a DFSAC base station (BS) is communicating with a set of $M$ UTs, while simultaneously sensing the targets in its surrounding environment. The BS has $M$ transmit antennas and $N$ receive antennas and each UT $m\in{\mathcal{M}}=\left\{1,\ldots,M\right\}$ has $K$ receive antennas. The reason why the number of UTs is set to the same value as that of the BS transmit antennas will be detailed in Section \ref{Reason}. Let $\mathbf{X}=\left[{\mathbf x}_1 \ldots {\mathbf x}_L\right]\in{\mathbbmss{C}}^{M\times L}$ be a DFSAC signal matrix, with $L$ being the length of the communication frame/sensing pulse. From a sensing perspective, ${\mathbf x}_l\in{\mathbbmss{C}}^{M\times1}$ for $l\in{\mathcal{L}}=\left\{1,\ldots,L\right\}$ represents the sensing snapshot transmitted at the $l$th time slot. For communications, ${\mathbf x}_l$ is the $l$th data symbol vector. {\color{blue}Under the framework of MIMO-ISAC, the signal matrix $\mathbf{X}$ can be written as
{\setlength\abovedisplayskip{2pt}
\setlength\belowdisplayskip{2pt}
\begin{align}\label{dual_function_signal_matrix}
{\mathbf{X}}={\mathbf{W}}{\mathbf{S}}={\mathbf{P}}{\bm\Xi}^{1/2}{\mathbf{S}},
\end{align}
}where ${\mathbf{W}}={\mathbf{P}}{\bm\Xi}^{1/2}\in{\mathbbmss{C}}^{M\times M}$ is the precoding matrix, ${\mathbf{P}}=\left[{\mathbf{p}}_1\ldots{\mathbf{p}}_M\right]\in{\mathbbmss{C}}^{M\times M}$ stores the normalized precoders with $\lVert{\mathbf{p}}_m\rVert^2=1$, $\forall m$, ${\bm\Xi}={\mathsf{diag}}\{{p_1},\ldots,{p_M}\}\succeq{\mathbf 0}$ is the power allocation matrix subject to the power budget $\sum_{m=1}^{M}p_m\leq p$}, and $\mathbf{S}=\left[{\mathbf{s}}_1\ldots{\mathbf{s}}_M\right]^{\mathsf{H}}\in{\mathbbmss{C}}^{M\times L}$ contains $M$ unit-power data streams intended for the $M$ UTs. Here, ${\mathbf{s}}_{m}^{\mathsf{H}}\in{\mathbbmss{C}}^{1\times L}$ and $\sqrt{p_m}{\mathbf{p}}_{m}\in{\mathbbmss{C}}^{M\times1}$ denote the data stream dedicated for UT $m$ and the associated precoding vector, respectively. The data streams are assumed to be independent with each other so that $L^{-1}{\mathbf{S}}{\mathbf{S}}^{\mathsf{H}}\approx \mathbf{I}_M$ \cite{Liu2022_TSP}.
\vspace{-10pt}
\subsection{Communication Model}\label{Comm_Model}
\vspace{-5pt}
The observation at each UT $m\in{\mathcal M}$ can be written as
{\setlength\abovedisplayskip{2pt}
\setlength\belowdisplayskip{2pt}
\begin{equation}
{\mathbf{Y}}_{{\rm{c}},m}={\mathbf{H}}_{m}(\sqrt{p_m}{\mathbf{p}}_{m}{\mathbf{s}}_{m}^{\mathsf{H}}+\sum\nolimits_{m'\neq m}\sqrt{p_{m'}}{\mathbf{p}}_{m'}{\mathbf{s}}_{m'}^{\mathsf{H}})+{\mathbf{N}}_{{\rm{c}},m},\nonumber
\end{equation}
}where ${\mathbf{H}}_{m}\in{\mathbbmss{C}}^{K\times M}$ is the communication channel matrix and ${\mathbf{N}}_{{\rm{c}},m}\in{\mathbbmss{C}}^{K\times L}$ is the additive white Gaussian noise (AWGN) matrix. It is assumed that all the communication links shown in {\figurename} {\ref{Figure1}} suffer Rayleigh fading. In this case, ${\mathbf{N}}_{{\rm{c}},m}$ and ${\mathbf{H}}_{m}$ are mutually independent complex Gaussian matrices, whose elements are independent and identically distributed (i.i.d.) with zero mean and unit variance. {\color{blue}We consider that each UT $m$ has access to its effective channel ${\mathbf{P}}_m={\mathbf{H}}_m{\mathbf{P}}\in{\mathbbmss{C}}^{K\times M}$, where ${\mathbf H}_m$ can be estimated by UT $m$ through downlink pilots and $\mathbf P$ can be fed to the UTs from the BS via an error-free link. As will be shown later, $\mathbf P$ is independent of the instantaneous channel realization $\{{\mathbf H}_m\}_{\forall m}$, which can be fixed for an extended period of time. Accordingly, the system overhead due to the feedback of $\mathbf P$ could be negligible. We further assume that the channel estimation error is negligible to characterize the performance bound. Imperfect channel acquisition can lead to extra interference and thus reduce the CR, whose influence will be detailed in our future works.}

After receiving ${\mathbf{Y}}_{{\rm{c}},m}$, UT $m$ adopts a normalized equalizer $\mathbf{v}_{m}\in{\mathbbmss{C}}^{K\times 1}$ to eliminate the inter-user interference (IUI). {\color{blue}By assuming $K\geq M$\footnote{{\color{blue}When $K<M$, the IUI cannot be thoroughly eliminated, thus yielding a reduced sum CR. Besides, in this case, maximizing the sum CR requires the joint design of power allocation and interference cancellation, which is generally NP-hard. This problem is still open and left for future work.}}, we have $\mathbf{v}_{m}=\frac{\mathbf{q}_{m}}{\lVert\mathbf{q}_{m}\rVert}$, where $\mathbf{q}_{m}$ denotes the $m$th column of matrix ${\mathbf{P}}_m\left({\mathbf{P}}_m^{\mathsf{H}}{\mathbf{P}}_m\right)^{-1}$. It is worth noting that the assumption of $K\geq M$ generally holds for some low-cost and low-power small cells where the BS has the same number of antennas as UTs, or even less.} Accordingly, the received SNR at UT $m$ is given as $\gamma_{m}=p_m\lvert{\mathbf{v}}_{m}^{\mathsf{H}}{\mathbf{H}}_{m}{\mathbf{p}}_{m}\rvert^2$. {\color{blue}In order to reduce system overhead caused by acquiring channel state information (CSI) at the BS, it is assumed that the BS does not have the global CSI $\{{\mathbf H}_m\in{\mathbbmss C}^{K\times M}\}_{\forall m}$ and only $\{{\mathbf{v}}_{m}^{\mathsf{H}}{\mathbf{H}}_{m}\in{\mathbbmss C}^{1\times M}\}_{\forall m}$\footnote{{\color{blue}We comment that the price of the overhead reduction mainly lies in the calculation of ${\mathbf v}_m$ at each UT $m$. In fact, the involved computational complexity is at the same order as that in designing a zero-forcing-based equalizer, which is acceptable for most communication terminals.}} is fed back to or estimated by the BS for power control. In this case, $\mathbf P$ is designed regardless of the global CSI $\{{\mathbf H}_m\}_{\forall m}$, and ${\bm\Xi}$ can be optimized to improve the CR. After optimizing $\bm\Xi$, the BS should feed $p_m$ to UT $m$ for decoding.} The sum CR reads $\mathcal{R}_{\rm{c}}=\sum\nolimits_{m=1}^{M}\log_2\left(1+\gamma_{m}\right)$.
\vspace{-10pt}
\subsection{Sensing Model}
\vspace{-5pt}
By transmitting $\mathbf{X}$ to sense the targets, the BS observes the following reflected echo signal matrix at its receiver: \cite{Liu2022_TSP,Tang2019_TSP,Ouyang2022_WCL}
{\setlength\abovedisplayskip{2pt}
\setlength\belowdisplayskip{2pt}
\begin{align}\label{reflected_echo_signal_matrix}
{\mathbf{Y}}_{\rm{s}}={\mathbf{G}}{\mathbf{X}}+{\mathbf{N}}_{\rm{s}},
\end{align}
}where ${\mathbf{N}}_{\rm{s}}\in{\mathbbmss{C}}^{N\times L}$ is the AWGN matrix with each entry having zero mean and unit variance, and $\mathbf{G}=\left[{\mathbf g}_1 \ldots {\mathbf{g}}_N\right]^{\mathsf{T}}\in{\mathbbmss{C}}^{N\times M}$ represents the target response matrix with ${\mathbf{g}}_n\in{\mathbbmss{C}}^{M\times 1}$ for $n\in{\mathcal{N}}=\left\{1,\ldots,N\right\}$ representing the target response from the transmit antenna array to the $n$th receive antenna. The target response matrix is modeled as \cite{Yuan2021_TVT,Liu2022_TSP,Tang2019_TSP,Ouyang2022_WCL}
{\setlength\abovedisplayskip{2pt}
\setlength\belowdisplayskip{2pt}
\begin{align}
\mathbf{G}=\sum\nolimits_{t}\beta_t{\mathbf{a}}\left(\theta_t\right){\mathbf{b}}^{\mathsf{H}}\left(\theta_t\right),
\end{align}
}where $\beta_t\sim{\mathcal{CN}}\left(0,\sigma_t^2\right)$ is the complex amplitude of the $t$th target with $\sigma_t^2$ representing the average strength, ${\mathbf{a}}\left(\theta_t\right)\in{\mathbbmss{C}}^{N\times1}$ and ${\mathbf{b}}\left(\theta_t\right)\in{\mathbbmss{C}}^{M\times1}$ are the associated receive and transmit array steering vectors, respectively, and $\theta_t$ is its direction of arrival. By considering that the receive antennas at the BS are widely separated, we have $\mathbf{g}_n\sim{\mathcal{CN}}\left({\mathbf{0}},\mathbf{R}\right)$ for $n\in{\mathcal{N}}$ and ${\mathbbmss{E}}\left\{{\mathbf{g}}_n{\mathbf{g}}_{n'}^{\mathsf{H}}\right\}={\mathbf{0}}$ for $n\neq n'$ \cite{Tang2019_TSP}.

Compared with the instantaneous target response ${\mathbf G}$, the correlation matrix ${\mathbf{R}}\in{\mathbbmss{C}}^{M\times M}$ are rather fixed for a longer period of time. It is not difficult for the BS to obtain $\mathbf R$ through long-term feedback. In the sequel, we assume that the BS has access to ${\mathbf R}$. This is a commonly used assumption in the literature, see, e.g., \cite{Tang2019_TSP}. Besides, we assume that none of the targets are registered communication UTs in the system. If the targets are also communication UTs, then we can exploit the sensed results to assist the communications, which is beyond the scope of this work.

Essentially, the aim of sensing is to extract environmental information contained in $\mathbf{G}$, e.g., the direction and reflection coefficient of each target, from the reflected echo signal ${\mathbf{Y}}_{\rm{s}}$ \cite{Yuan2021_TVT,Liu2022_TSP,Tang2019_TSP,Ouyang2022_WCL}. Particularly, the mutual information (MI) between ${\mathbf{Y}}_{\rm{s}}$ and $\mathbf{G}$ conditioned on the DFSAC signal ${\mathbf{X}}$ characterizes the information-theoretic limits on how much environmental information can be extracted, which is also referred to as the sensing MI \cite{Tang2019_TSP}. On this basis, from an information-theoretic perspective, we adopt the SR as the performance metric of sensing, which is defined as the sensing MI per unit time \cite{Zhang2021_JSTSP,Yuan2021_TVT,Ouyang2022_WCL}. Assuming that each DFSAC symbol lasts 1 unit time, we write the SR as ${\mathcal{R}}_{\rm{s}}=L^{-1}I\left({\mathbf{Y}}_{\rm{s}};\mathbf{G}|\mathbf{X}\right)$, where $I\left(X;Y|Z\right)$ denotes the MI between $X$ and $Y$ conditioned on $Z$.

Given the MIMO-ISAC framework, we intend to analyze its S\&C performance by investigating the CR $\mathcal{R}_{\rm{c}}$ and SR $\mathcal{R}_{\rm{s}}$. Note that both $\mathcal{R}_{\rm{c}}$ and $\mathcal{R}_{\rm{s}}$ are influenced by the precoding matrix $\mathbf{W}$. Yet, it is challenging to find a $\mathbf{W}$ that can maximize $\mathcal{R}_{\rm{s}}$ and $\mathcal{R}_{\rm{c}}$ simultaneously. As a compromise, we consider three typical scenarios to unveil further system insights. The first scenario is the S-C design that aims to maximize the SR, the second scenario is the C-C design that aims to maximize the CR, and the third scenario is the Pareto optimal design that aims to find the Pareto boundary of the rate region.
\vspace{-10pt}
\section{Performance of MIMO-ISAC}
\subsection{Sensing-Centric Design}\label{Reason}
\subsubsection{Performance of Sensing}
Under the S-C design, the precoding matrix $\mathbf{W}$ is set to maximize ${\mathcal{R}}_{\rm{s}}$. Particularly, the SR can be calculated as \cite{Tang2019_TSP}
{\setlength\abovedisplayskip{2pt}
\setlength\belowdisplayskip{2pt}
\begin{align}
{\mathcal{R}}_{\rm{s}}=L^{-1}N\log_2\det\left({\mathbf{I}}_M+L{\mathbf{W}}^{\mathsf{H}}{\mathbf{R}}{\mathbf{W}}\right).
\end{align}
}Under the S-C design, the precoding matrix satisfies
{\setlength\abovedisplayskip{2pt}
\setlength\belowdisplayskip{2pt}
\begin{align}
\mathbf{W}_{\rm{s}}=\arg\max\nolimits_{{\mathsf{tr}}\left({\mathbf{W}}{\mathbf{W}}^{\mathsf{H}}\right)\leq p}{\log_2\det\left({\mathbf{I}}_M+L{\mathbf{W}}^{\mathsf{H}}{\mathbf{R}}{\mathbf{W}}\right)}.
\end{align}
}For analytical tractability, we assume that ${\mathbf{R}}\succ{\mathbf{0}}$\footnote{When ${\mathbf{R}}\succ{\mathbf{0}}$, the rank of $\mathbf G$ is given by $M$. By \cite{Liu2022_TSP}, to recover the rank-$M$ matrix $\mathbf G$, we should transmit at least $M$ independent signal streams. For brevity, we assume that the number of data streams or UTs equals $M$. Actually, the derived results in this letter can be extended to the case with more than $M$ UTs, which is left as a potential direction for future work.}. Then, the following theorem provides an exact expression for the SR as well as its high-SNR approximation.
\vspace{-5pt}
\begin{theorem}\label{Sensing_Rate_S_C_Theorem}
In the S-C design, the maximum SR is given by
{\setlength\abovedisplayskip{2pt}
\setlength\belowdisplayskip{2pt}
\begin{align}
\mathcal{R}_{\rm{s}}^{\rm{s}}={N}{L^{-1}}\sum\nolimits_{m=1}^{M}\log_2\left(1+L\lambda_ms_m^{\star}\right),
\end{align}
}where $\left\{\lambda_m>0\right\}_{m=1}^{M}$ are the eigenvalues of matrix ${\mathbf{R}}$ and $s_{m}^{\star}=\max\left\{0,\frac{1}{\nu}-\frac{1}{L\lambda_m}\right\}$ with
{\setlength\abovedisplayskip{2pt}
\setlength\belowdisplayskip{2pt}
\begin{align}
\sum_{m=1}^{M}\max\left\{0,\frac{1}{\nu}-\frac{1}{L\lambda_m}\right\}=p.\nonumber
\end{align}
}The maximum SR is attained when ${\mathbf{W}}_{\rm{s}}={\mathbf{U}}{\bm\Delta}_{\rm{s}}^{1/2}$, where ${\mathbf{U}}{\mathsf{diag}}\left\{\lambda_1,\ldots,\lambda_{M}\right\}{\mathbf{U}}^{\mathsf{H}}$ denotes the eigendecomposition (ED) of ${\mathbf{R}}$ and ${\bm\Delta}_{\rm{s}}={\mathsf{diag}}\left\{s_1^{\star},\ldots,s_M^{\star}\right\}$. As $p\rightarrow\infty$,
{\setlength\abovedisplayskip{2pt}
\setlength\belowdisplayskip{2pt}
\begin{align}\label{Sensing_Rate_S_C_Asymptotic}
\mathcal{R}_{\rm{s}}^{\rm{s}}
\approx\frac{NM}{L}\left(\log_2{p}+\frac{1}{M}\sum\nolimits_{m=1}^{M}\log_2\left(\frac{L\lambda_m}{M}\right)\right).
\end{align}
}\end{theorem}
\vspace{-5pt}
\begin{IEEEproof}
Similar to the proof of \cite[Theorem 3]{Ouyang2022_WCL}.
\end{IEEEproof}
\vspace{-5pt}
\begin{remark}
The results in \eqref{Sensing_Rate_S_C_Asymptotic} suggest that the high-SNR slope of the SR under the S-C design is given by $\frac{NM}{L}$.
\end{remark}
\vspace{-5pt}
\subsubsection{Performance of Communications}
Turn now to the communication performance. Particularly, the CR of UT $m$ is given by ${\overline{\mathcal{R}}}_{{\rm{c}},m}^{\rm{s}}=\log_2\left(1+s_m^{\star}\rho_m\right)$ with $\rho_m=\lvert{\mathbf{v}}_{m}^{\mathsf{H}}{\mathbf{h}}_{m}\rvert^2$ and ${\mathbf{h}}_{m}\in{\mathbbmss{C}}^{K\times1}$ being the $m$th column of matrix ${\mathbf{H}}_m{\mathbf{U}}\in{\mathbbmss{C}}^{K\times M}$. Specifically, we utilize the OP and ergodic CR (ECR) to evaluate the communication performance. The following theorem provides an exact expression for the sum ECR ${{\mathcal{R}}}_{{\rm{c}}}^{\rm{s}}={\mathbbmss{E}}\{{\overline{\mathcal{R}}}_{{\rm{c}}}^{\rm{s}}\}$ with ${\overline{\mathcal{R}}}_{{\rm{c}}}^{\rm{s}}=\sum_{m=1}^{M}{\overline{\mathcal{R}}}_{{\rm{c}},m}^{\rm{s}}$ as well as its high-SNR approximation.
\vspace{-5pt}
\begin{theorem}\label{Ergodic_Communication_Rate_S_C_Theorem}
In the S-C design, the sum ECR is given by
{\setlength\abovedisplayskip{2pt}
\setlength\belowdisplayskip{2pt}
\begin{align}
{{\mathcal{R}}}_{{\rm{c}}}^{\rm{s}}&=\sum\nolimits_{m=1}^{M}\sum\nolimits_{\mu=0}^{K'}\frac{(-1/s_{m}^{\star})^{K'-\mu}}{(K'-\mu)!\ln{2}}
\left(-{\rm{e}}^{-\frac{1}{s_{m}^{\star}}}{\rm{Ei}}\left(\frac{1}{s_{m}^{\star}}\right)\right.\nonumber\\
&+\left.\sum\nolimits_{i=1}^{K'-\mu}(i-1)!\left(-1/s_{m}^{\star}\right)^{-i}\right),\label{Ergodic_Communication_Rate_S_C_Basic}
\end{align}
}where $K'=K-M$ and ${\rm{Ei}}(x)=-\int_{-x}^{\infty}{\rm{e}}^{-t}t^{-1}{\rm{d}}t$ is the exponential integral function \cite[Eq. (8.211.1)]{Ryzhik2007}. As $p\rightarrow\infty$,
{\setlength\abovedisplayskip{2pt}
\setlength\belowdisplayskip{2pt}
\begin{align}\label{Ergodic_Communication_Rate_S_C_Asymptotic}
{{\mathcal{R}}}_{{\rm{c}}}^{\rm{s}}
\approx M(\log_2{p}-\log_2{M}+\psi\left(K'+1\right)/\ln{2}),
\end{align}
}where $\psi\left(x\right)=\frac{{\rm d}}{{\rm d}x}\ln{\Gamma\left(x\right)}$ is the Digamma function \cite[Eq. (6.461)]{Ryzhik2007} and $\Gamma\left(x\right)=\int_{0}^{\infty}t^{x-1}{\rm e}^{-t}{\rm d}t$ is the gamma function \cite[Eq. (6.1.1)]{Ryzhik2007}.
\end{theorem}
\vspace{-5pt}
\begin{IEEEproof}
Please refer to Appendix \ref{Proof_Ergodic_Communication_Rate_S_C_Theorem} for more details.
\end{IEEEproof}
\vspace{-5pt}
\begin{remark}
The results in \eqref{Ergodic_Communication_Rate_S_C_Asymptotic} suggest that the high-SNR slope of the sum ECR under the S-C design is given by $M$.
\end{remark}
\vspace{-5pt}
It is challenging to derive a closed-form expression for the OP, i.e., ${{\mathcal{P}}}_{{\rm{c}}}^{\rm{s}}=\Pr({\overline{\mathcal{R}}}_{{\rm{c}}}^{\rm{s}}<\mathcal{R}_0)$ with $\mathcal{R}_0$ denoting the target sum CR. Thus, we focus more on its high-SNR properties.
\vspace{-5pt}
\begin{theorem}\label{Outage_Probability_UT_S_C_Theorem}
As $p\rightarrow\infty$, the OP of the sum CR achieved by the S-C design satisfies ${{\mathcal{P}}}_{{\rm{c}}}^{\rm{s}}\simeq{\mathcal{O}}\left(p^{-M(K+M-1)}\right)$. The notation $f(x)={\mathcal{O}}\left(g(x)\right)$ means that $\limsup_{x\rightarrow\infty}\frac{\left|f(x)\right|}{g(x)}<\infty$.
\end{theorem}
\vspace{-5pt}
\begin{IEEEproof}
Please refer to Appendix \ref{Proof_Outage_Probability_UT_S_C_Theorem} for more details.
\end{IEEEproof}
\vspace{-5pt}
\begin{remark}
The above results suggest that a diversity of $M(K-M+1)$ is achievable for the OP under the S-C design.
\end{remark}
\vspace{-15pt}
\subsection{Communications-Centric Design}
Having investigated the S\&C performance under the S-C design, we now move to the C-C design.
\subsubsection{Performance of Communications}
It is worth noting that $\mathbf{H}_m$ has the same statistical properties as $\mathbf{H}_m{\mathbf{U}}$. In light of this fact as well as the conclusion drawn in Theorem \ref{Sensing_Rate_S_C_Theorem}, we design the C-C precoding matrix as ${\mathbf{W}}={\mathbf{P}}{\bm\Delta}_{\rm{c}}^{1/2}$ with ${\mathbf P}={\mathbf U}$ and ${\bm\Delta}_{\rm{c}}={\mathsf{diag}}\left\{c_1,\ldots,c_{M}\right\}$, where $\sum_{m=1}^{M}c_{m}\leq p$ and $c_m\geq0$ for $m\in{\mathcal{M}}$. The resulting precoding matrix satisfies
{\setlength\abovedisplayskip{2pt}
\setlength\belowdisplayskip{2pt}
\begin{align}
{\mathbf{W}}_{\rm{c}}={\arg\max}_{{\mathbf{W}}={\mathbf{U}}{\bm\Delta}_{\rm{c}}^{1/2}}\sum\nolimits_{m=1}^{M}\log_2\left(1+c_m\rho_m\right),
\end{align}
}where $\log_2\left(1+c_m\rho_m\right)$ calculates the CR of UT $m$. Particularly, the following lemma provides an expression for the maximum sum CR achieved by ${\mathbf{W}}_{\rm{c}}$.
\vspace{-5pt}
\begin{lemma}\label{Communication_Rate_C_C_Lemma}
Under the C-C design, the maximum sum CR is
{\setlength\abovedisplayskip{2pt}
\setlength\belowdisplayskip{2pt}
\begin{align}
{\overline{\mathcal{R}}}_{\rm{c}}^{\rm{c}}=\sum\nolimits_{m=1}^{M}\log_2\left(1+\rho_mc_m^{\star}\right),
\end{align}
}where $c_{m}^{\star}=\max\left\{0,\frac{1}{\upsilon}-\frac{1}{\rho_m}\right\}$ with $\sum_{m=1}^{M}c_{m}^{\star}=p$. The maximum sum CR is attained when $c_m=c_m^{\star}$ for $m\in{\mathcal{M}}$.
\end{lemma}
\vspace{-5pt}
\begin{IEEEproof}
This lemma can be directly proved by using the water-filling procedure \cite{Heath2018}.
\end{IEEEproof}
We note that deriving a closed-form expression of the sum ECR ${{\mathcal{R}}}_{\rm{c}}^{\rm{c}}={\mathbbmss{E}}\{{\overline{\mathcal{R}}}_{\rm{c}}^{\rm{c}}\}$ is a hard task. To unveil more insights, we characterize its high-SNR behaviour in Theorem \ref{Ergodic_Communication_Rate_C_C_Theorem}.
\vspace{-5pt}
\begin{theorem}\label{Ergodic_Communication_Rate_C_C_Theorem}
As $p\rightarrow\infty$, ${{\mathcal{R}}}_{\rm{c}}^{\rm{c}}$ satisfies
{\setlength\abovedisplayskip{2pt}
\setlength\belowdisplayskip{2pt}
\begin{align}\label{Ergodic_Communication_Rate_C_C_Asymptotic}
{{\mathcal{R}}}_{{\rm{c}}}^{\rm{c}}
\approx M(\log_2{p}-\log_2{M}+\psi\left(K'+1\right)/\ln{2}),
\end{align}
}\end{theorem}
\vspace{-5pt}
\begin{IEEEproof}
Please refer to Appendix \ref{Proof_Ergodic_Communication_Rate_C_C_Theorem} for more details.
\end{IEEEproof}
\vspace{-5pt}
\begin{remark}
The results in \eqref{Ergodic_Communication_Rate_C_C_Asymptotic} suggest that the high-SNR slope of the sum ECR under the C-C design is given by $M$.
\end{remark}
\vspace{-5pt}
\vspace{-5pt}
\begin{remark}\label{CR_Comparision}
By comparing \eqref{Ergodic_Communication_Rate_S_C_Asymptotic} and \eqref{Ergodic_Communication_Rate_C_C_Asymptotic}, we observe that the CR achieved by the C-C design has the same asymptotic behaviour as that achieved by the S-C design.
\end{remark}
\vspace{-5pt}
Turn to the OP ${{\mathcal{P}}}_{{\rm{c}}}^{\rm{c}}=\Pr({\overline{\mathcal{R}}}_{\rm{c}}^{\rm{c}}<\mathcal{R}_0)$, whose asymptotic behaviour is characterized as follows.
\vspace{-5pt}
\begin{theorem}\label{Outage_Probability_C_C_Theorem}
As $p\rightarrow\infty$, the OP of the sum CR achieved by the C-C design satisfies ${{\mathcal{P}}}_{{\rm{c}}}^{\rm{c}}\simeq{\mathcal{O}}\left(p^{-M(K+M-1)}\right)$.
\end{theorem}
\vspace{-5pt}
\begin{IEEEproof}
Similar to the proof of Theorem \ref{Outage_Probability_UT_S_C_Theorem}.
\end{IEEEproof}
\vspace{-5pt}
\begin{remark}
The above results suggest that a diversity order of $M(K-M+1)$ is achievable for the OP under the C-C design, which is the same as that achieved by the S-C design.
\end{remark}
\vspace{-5pt}
\subsubsection{Performance of Sensing}
For ${\mathbf W}={\mathbf{W}}_{\rm{c}}$, the SR reads
{\setlength\abovedisplayskip{2pt}
\setlength\belowdisplayskip{2pt}
\begin{align}
{\overline{\mathcal R}}_{\rm{s}}^{\rm{c}}={N}{L^{-1}}\sum\nolimits_{m=1}^{M}\log_2\left(1+L\lambda_mc_m^{\star}\right).
\end{align}
}Due to the statistics of $c_m^{\star}$, we define the average SR as ${{\mathcal R}}_{\rm{s}}^{\rm{c}}={\mathbbmss E}\{{\overline{\mathcal R}}_{\rm{s}}^{\rm{c}}\}$, which can be evaluated numerically. Besides, Theorem \ref{Average_Sensing_Rate_C_C_Theorem} describes the high-SNR behaviour of ${{\mathcal R}}_{\rm{s}}^{\rm{c}}$.
\vspace{-5pt}
\begin{theorem}\label{Average_Sensing_Rate_C_C_Theorem}
As $p\rightarrow\infty$, ${{\mathcal R}}_{\rm{s}}^{\rm{c}}$ satisfies
{\setlength\abovedisplayskip{2pt}
\setlength\belowdisplayskip{2pt}
\begin{align}\label{Average_Sensing_Rate_C_C_Asymptotic}
{{\mathcal{R}}}_{{\rm{s}}}^{\rm{c}}
\approx\frac{NM}{L}\left(\log_2{p}+\frac{1}{M}\sum\nolimits_{m=1}^{M}\log_2\left(\frac{L\lambda_m}{M}\right)\right).
\end{align}
}\end{theorem}
\vspace{-5pt}
\begin{IEEEproof}
Similar to the proof of Theorem \ref{Ergodic_Communication_Rate_C_C_Theorem}.
\end{IEEEproof}
\vspace{-5pt}
\begin{remark}\label{SR_Comparision}
The SR achieved by the C-C design involves the same asymptotic behaviour as that achieved by the S-C design.
\end{remark}
\vspace{-5pt}
Taken the conclusions in Remarks \ref{CR_Comparision} and \ref{SR_Comparision} together, we find that the C-C design degenerates to the S-C design in the high-SNR regime and vice versa.
\vspace{-10pt}
\subsection{Pareto Optimal Design}
In practice, the precoding matrix $\mathbf{W}$ can be designed to satisfy different qualities of services, which results in a communication-sensing performance tradeoff. To evaluate this tradeoff, we resort to the Pareto boundary of the SR-CR region. The Pareto boundary consists of SR-CR tuples at which it is impossible to improve one of the two rates without simultaneously decreasing the other \cite{Zhang2010_TSP}. More precisely, let $(\hat{\mathcal{R}}_{\rm{s}},\hat{\mathcal{R}}_{\rm{c}})$ denote a rate-tuple on the Pareto boundary, then there is no other rate-tuple $(\hat{\mathcal{R}}_{\rm{s}}^{\prime},\hat{\mathcal{R}}_{\rm{c}}^{\prime})$ with $\hat{\mathcal{R}}_{\rm{s}}^{\prime}\geq\hat{\mathcal{R}}_{\rm{s}}$, $\hat{\mathcal{R}}_{\rm{c}}^{\prime}\geq\hat{\mathcal{R}}_{\rm{c}}$, and $(\hat{\mathcal{R}}_{\rm{s}}^{\prime},\hat{\mathcal{R}}_{\rm{c}}^{\prime})\neq(\hat{\mathcal{R}}_{\rm{s}},\hat{\mathcal{R}}_{\rm{c}})$ \cite{Zhang2010_TSP}. Particularly, we design the precoding matrix as $\mathbf{W}_{\rm{p}}={\mathbf P}{\mathsf{diag}}\left\{\sqrt{p_1},\ldots,\sqrt{p_M}\right\}$ with ${\mathbf P}={\mathbf U}$, $\sum_{m=1}^{M}p_m\leq p$, and $p_m\geq0$, $\forall m$. By \cite{Zhang2010_TSP}, any rate-tuple on the Pareto boundary can be obtained via the rate-profile based method, i.e., solving the following problem:
{\setlength\abovedisplayskip{2pt}
\setlength\belowdisplayskip{2pt}
\begin{equation}\label{Problem_CR_SR_Tradeoff}
\max\nolimits_{{\mathbf{p}},\mathcal{R}}{\mathcal{R}},~{\rm{s.t.}}{\mathcal{R}}_{\rm{s}}\geq \alpha{\mathcal{R}},{\mathcal{R}}_{\rm{c}}\geq \bar{\alpha}{\mathcal{R}},{\mathbf{1}}^{\mathsf{T}}{\mathbf{p}}\!\leq\! p,p_m\!\geq\!0,
\end{equation}
}where $\alpha\in[0,1]$ is a particular rate-profile parameter, $\bar{\alpha}=1-\alpha$, and ${\mathbf{p}}=[p_1,\ldots,p_M]^{\mathsf{T}}$. We comment that problem \eqref{Problem_CR_SR_Tradeoff} is not equivalent to the weighted sum rate maximization problem (WSRMP) defined in \cite{Yuan2021_TVT}. Besides, solving the WSRMP cannot guarantee the finding of all Pareto-boundary points \cite{Zhang2010_TSP}. Problem \eqref{Problem_CR_SR_Tradeoff} is convex and can be solved via standard convex problem solvers such as CVX. For a given $\alpha$, let ${{\mathcal{R}}}_{\rm{c}}^{\alpha}$ and ${{\mathcal{R}}}_{\rm{s}}^{\alpha}$ denote the sum ECR and average SR achieved by the corresponding optimal precoding matrix, respectively. It follows that ${{\mathcal{R}}}_{\rm{c}}^{\alpha}\in\left[{{\mathcal{R}}}_{\rm{s}}^{\rm{c}},{{\mathcal{R}}}_{\rm{c}}^{\rm{c}}\right]$ and ${{\mathcal{R}}}_{\rm{s}}^{\alpha}\in\left[{{\mathcal{R}}}_{\rm{c}}^{\rm{s}},{{\mathcal{R}}}_{\rm{s}}^{\rm{s}}\right]$ with ${{\mathcal{R}}}_{\rm{s}}^{1}={{\mathcal{R}}}_{\rm{s}}^{\rm{s}}$ and ${{\mathcal{R}}}_{\rm{c}}^{0}={{\mathcal{R}}}_{\rm{c}}^{\rm{c}}$. Accordingly, we get the following corollaries.
\vspace{-5pt}
\begin{corollary}\label{CR_SR_Pareto_Boundary_Asymptotic}
For a sufficiently larger SNR, ${{\mathcal{R}}}_{\rm{s}}^{\alpha}\approx\frac{NM}{L}\left(\log_2{p}+\frac{1}{M}\sum\nolimits_{m=1}^{M}\log_2\left(\frac{L\lambda_m}{M}\right)\right)$ and ${{\mathcal{R}}}_{\rm{c}}^{\alpha}\approx M(\log_2{p}-\log_2{M}+\psi\left(K'+1\right)/\ln{2})$.
\end{corollary}
\vspace{-5pt}
\begin{IEEEproof}
This corollary can be proved by using the results in Theorems \ref{Ergodic_Communication_Rate_S_C_Theorem} and \ref{Ergodic_Communication_Rate_C_C_Theorem} as well as the Sandwich theorem.
\end{IEEEproof}
\vspace{-5pt}
\begin{corollary}\label{Outage_Probability_Pareto_Boundary_Asymptotic}
Let ${\overline{\mathcal R}}_{\rm{c}}^{\alpha}$ denote the sum CR for a given $\alpha$. Then, it has $\lim_{p\rightarrow\infty}\Pr({\overline{\mathcal R}}_{\rm{c}}^{\alpha}<\mathcal{R}_0)\simeq{\mathcal{O}}\left(p^{-M(K+M-1)}\right)$.
\end{corollary}
\vspace{-5pt}
\begin{IEEEproof}
Similar to the proof of Corollary \ref{CR_SR_Pareto_Boundary_Asymptotic}.
\end{IEEEproof}
\vspace{-5pt}
\begin{remark}
Any SR-CR tuple on the Pareto boundary has the same asymptotic behaviour in the high-SNR regime.
\end{remark}
\vspace{-5pt}
Let ${\mathcal{R}}^{\rm{s}}$ and ${\mathcal{R}}^{\rm{c}}$ denote the achievable SR and CR, respectively. Then, the rate region achieved by ISAC is given by
{\setlength\abovedisplayskip{2pt}
\setlength\belowdisplayskip{2pt}
\begin{align}
\mathcal{C}_{\rm{i}}=\left\{\left({\mathcal{R}}^{\rm{s}},{\mathcal{R}}^{\rm{c}}\right)|{\mathcal{R}}^{\rm{s}}\!\in\!\left[0,\mathcal{R}_{\rm{s}}^{\alpha}\right],
{\mathcal{R}}^{\rm{c}}\!\in\!\left[0,\mathcal{R}_{\rm{c}}^{\alpha}\right],\alpha\!\in\!\left[0,\!1\right]\right\}.\label{Rate_Regio_ISAC}
\end{align}
}\vspace{-20pt}
\begin{remark}
{\color{blue}In the above three scenarios, we set ${\mathbf P}={\mathbf U}$, which is solely determined by the second order statistics of the target response $\mathbf G$. As stated before, $\mathbf G$ or $\mathbf U$ can be fixed for an extended period of time, which is consistent with the statements in Section \ref{Comm_Model}.}
\end{remark}
\vspace{-15pt}
\section{Performance of MIMO-FDSAC}
We consider FDSAC as a baseline scenario, where the total bandwidth is partitioned into two sub-bands, one for sensing only and the other for communications. Besides, the total power is also partitioned into two parts for sensing and communications, respectively. Specifically, we assume $\kappa$ fraction of the total bandwidth and $\mu$ fraction of the total power is used for communications. Based on \cite{Ouyang2022_WCL}, the sum CR and the SR are given by $\overline{\mathcal{R}}_{\rm{c}}^{\rm{f}}=\max_{\sum_{m=1}^{M}a_m\leq \mu p}\sum_{m=1}^{M}\kappa\log_2(1+\frac{a_m}{\kappa}\rho_m)$ and $\mathcal{R}_{\rm{s}}^{\rm{f}}=\frac{N(1-\kappa)}{L}\max_{\sum_{m=1}^{M}b_m\leq (1-\mu)p}\sum\nolimits_{m=1}^{M}\log_2\left(1+\frac{L\lambda_m}{1-\kappa}b_m\right)$, respectively. Accordingly, we derive the following corollary.
\vspace{-5pt}
\begin{corollary}\label{FDSAC_Diversity_Order}
As $p\rightarrow\infty$, the OP of the sum CR achieved by FDSAC satisfies $\Pr(\overline{\mathcal{R}}_{\rm{c}}^{\rm{f}}<\mathcal{R}_0)\simeq{\mathcal{O}}\left(p^{-M(K+M-1)}\right)$.
\end{corollary}
\vspace{-5pt}
\begin{IEEEproof}
Similar to the proof of Corollary \ref{CR_SR_Pareto_Boundary_Asymptotic}.
\end{IEEEproof}
\vspace{-5pt}
\begin{corollary}\label{FDSAC_HSPO}
The high-SNR slopes of $\mathcal{R}_{\rm{c}}^{\rm{f}}={\mathbbmss{E}}\{\overline{\mathcal{R}}_{\rm{c}}^{\rm{f}}\}$ and $\mathcal{R}_{\rm{s}}^{\rm{f}}$ are given by $\kappa M$ and $(1-\kappa)\frac{NM}{L}$, respectively.
\end{corollary}
\vspace{-5pt}
\begin{IEEEproof}
Similar to the proofs of Theorems \ref{Sensing_Rate_S_C_Theorem} and \ref{Ergodic_Communication_Rate_C_C_Theorem}.
\end{IEEEproof}
Moreover, the rate region achieved by FDSAC is given by
{\setlength\abovedisplayskip{2pt}
\setlength\belowdisplayskip{2pt}
\begin{align}
\mathcal{C}_{\rm{f}}=\left\{\left({\mathcal{R}}^{\rm{s}},{\mathcal{R}}^{\rm{c}}\right)\left|
\begin{aligned}
&{\mathcal{R}}^{\rm{s}}\in\left[0,\mathcal{R}_{\rm{s}}^{\rm{f}}\right],
{\mathcal{R}}^{\rm{c}}\in\left[0,\mathcal{R}_{\rm{c}}^{\rm{f}}\right],\\
&\kappa\in\left[0,1\right],\mu\in\left[0,1\right]
\end{aligned}
\right.\right\}.\label{Rate_Regio_FDSAC}
\end{align}
}After completing all the analyses, we summarize the results related to diversity order and high-SNR slope in Table \ref{table1}.
\vspace{-5pt}
\begin{remark}\label{High_SNR_Slope_Compare}
The results in Table \ref{table1} suggest that ISAC and FDSAC yield the same diversity order in terms of the sum CR. Moreover, since $\kappa\in[0,1]$, we note that ISAC yields larger high-SNR slopes than FDSAC, which means that ISAC provides more degrees of freedom than FDSAC in terms of both communications and sensing \cite{Heath2018}.
\end{remark}
\vspace{-5pt}
We then compare the rate regions $\mathcal{C}_{\rm{i}}$ and $\mathcal{C}_{\rm{f}}$ as follows.
\vspace{-5pt}
\begin{theorem}\label{theorem_Rate_Region_Comparision}
The achievable rate regions satisfy $\mathcal{C}_{\rm{f}}\subseteq\mathcal{C}_{\rm{i}}$.
\end{theorem}
\vspace{-5pt}
\begin{IEEEproof}
Please refer to Appendix \ref{Proof_theorem_Rate_Region_Comparision} for more details.
\end{IEEEproof}
\vspace{-5pt}
\begin{remark}
The results in Theorem \ref{theorem_Rate_Region_Comparision} suggest that the rate region achieved by FDSAC is entirely covered by that achieved by ISAC. This superiority mainly originates from ISAC's integrated utilization of spectrum and power resources.
\end{remark}
\vspace{-5pt}

\begin{table}[!t]
\centering
\setlength{\abovecaptionskip}{0pt}
\resizebox{0.6\textwidth}{!}{
\begin{tabular}{|c|cc|c|}
\hline
\multirow{2}{*}{System} & \multicolumn{2}{c|}{Sum CR}                        & SR                       \\ \cline{2-4}
                        & \multicolumn{1}{c|}{$\mathcal{D}$} & $\mathcal{S}$ & $\mathcal{S}$            \\ \hline
ISAC                    & \multicolumn{1}{c|}{$M(K+M-1)$}    & $M$           & ${NM}/{L}$           \\ \hline
FDSAC                   & \multicolumn{1}{c|}{$M(K+M-1)$}    & $\kappa M$    & $(1-\kappa){NM}/{L}$ \\ \hline
\end{tabular}}
\caption{Diversity Order ($\mathcal{D}$) and High-SNR Slope ($\mathcal{S}$)}
\label{table1}
\end{table}
\vspace{-10pt}
\section{Numerical Results}
Simulation results will be presented to evaluate the S\&C performance of ISAC systems and also verify the accuracy of the developed analytical results. The parameters used for simulation are listed as follows: $M=4$, $N=5$, $K=4$, $L=30$, and the eigenvalues of $\mathbf{R}$ are $\{1,0.1,0.05,0.01\}$.

\begin{figure}[!t]
    \centering
    \subfigbottomskip=0pt
	\subfigcapskip=-5pt
\setlength{\abovecaptionskip}{0pt}
    \subfigure[OP of the sum CR.]
    {
        \includegraphics[height=0.3\textwidth]{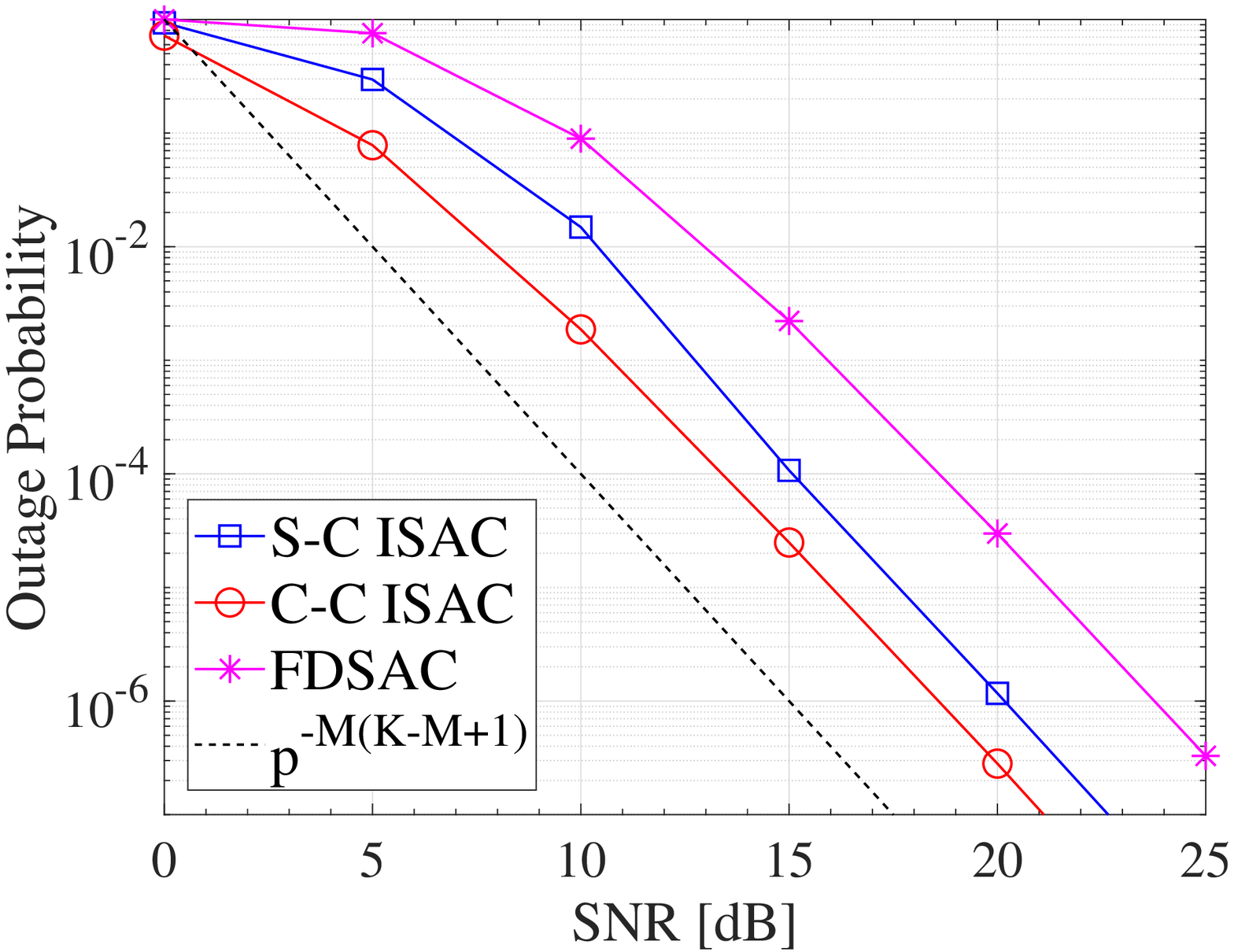}
	   \label{fig2a}	
    }
   \subfigure[Sum ECR.]
    {
        \includegraphics[height=0.3\textwidth]{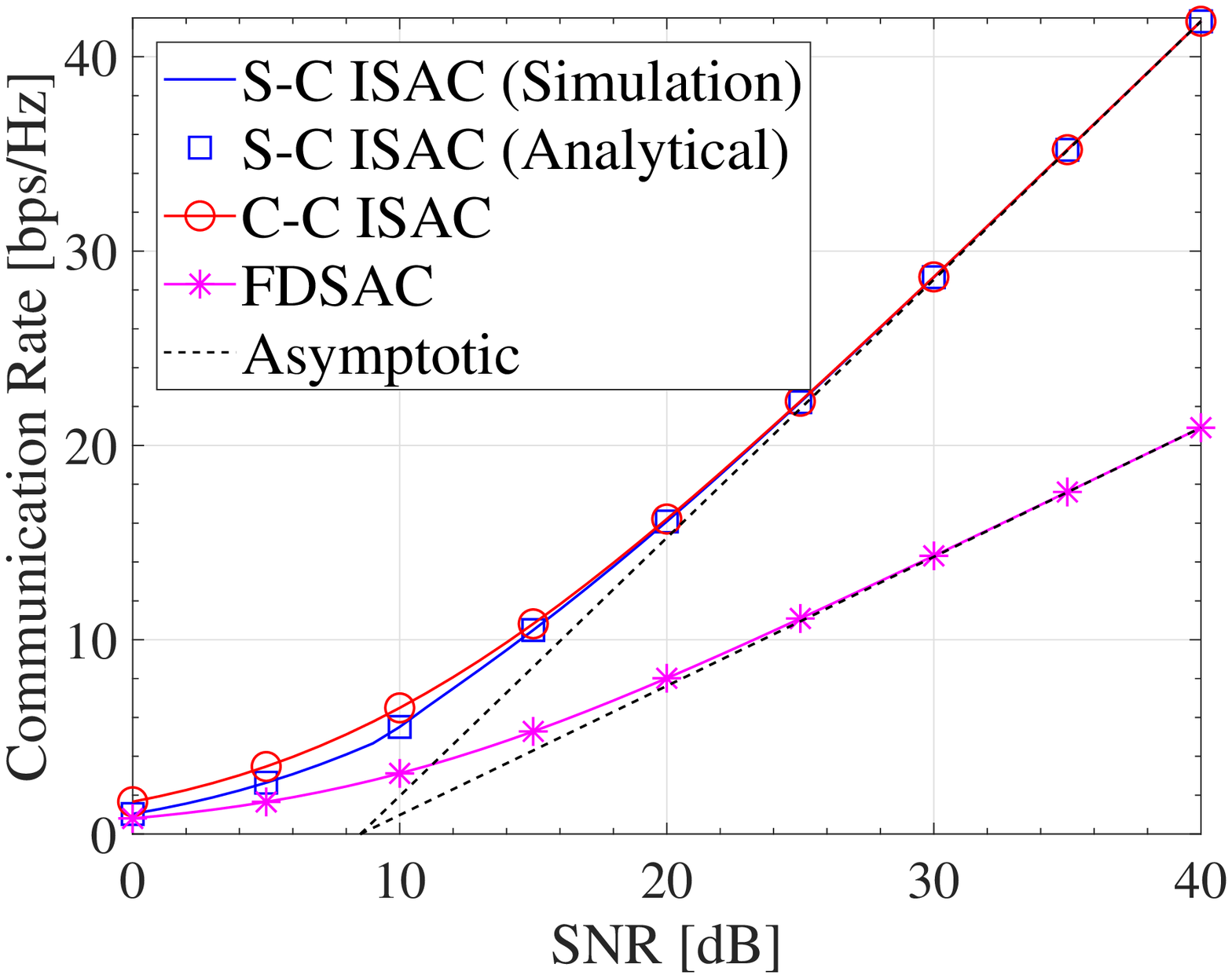}
	   \label{fig2b}	
    }
\caption{Performance of communications. ${\mathcal{R}}_0=2$ bps/Hz.}
    \label{Figure2}
    \vspace{-20pt}
\end{figure}

%\begin{figure}[!t]
%\centering
%\setlength{\abovecaptionskip}{0pt}
%\includegraphics[height=0.182\textwidth]{SR.eps}
%\caption{Performance of sensing. $\kappa=\mu=0.5$.}
%\label{Figure3}
%\vspace{-20pt}
%\end{figure}

{\figurename} {\ref{fig2a}} and {\figurename} {\ref{fig2b}} plot the OP and sum ECR versus the SNR $p$ for $\kappa=\mu=0.5$, respectively. As {\figurename} {\ref{fig2a}} shows, C-C ISAC achieves the lowest OP while FDSAC achieves the highest OP. In the high-SNR regime, the OP curves for all the presented cases are parallel to the one representing $p^{-M(K-M+1)}$, which suggests that the achievable diversity order obtained in the previous section is tight. This also suggests that ISAC yields the same diversity order as FDSAC. Let us now turn to {\figurename} {\ref{fig2b}}. As expected, C-C ISAC attains the best ECR performance among the three presented cases. Besides, the analytical results fit well with the simulations, and the asymptotic results accurately track the provided simulation results in the high-SNR regime. Particularly, it can be seen from {\figurename} {\ref{fig2b}} that ISAC achieves a larger high-SNR slope than FDSAC and the ECRs achieved by C-C ISAC and S-C ISAC have the same high-SNR asymptotic behaviour, which is consistent with the results shown in Remarks \ref{CR_Comparision} and \ref{High_SNR_Slope_Compare}.

In {\figurename} {\ref{Figure3}}, the SR is shown as a function of the SNR $p$. It can be seen from this graph that S-C ISAC is capable of achieving the best SR performance among the presented four cases. Moreover, as {\figurename} {\ref{Figure3}} shows, the asymptotic results track the provided simulation results accurately in the high-SNR regime. From the data in {\figurename} {\ref{Figure3}}, it is apparent that ISAC achieves a larger high-SNR slope than FDSAC and S-C ISAC and C-C ISAC yield the same high-SNR slope, which agrees with the conclusions in Remark \ref{SR_Comparision} and Remark \ref{High_SNR_Slope_Compare}.

%\begin{figure}[!t]
%\centering
%\setlength{\abovecaptionskip}{0pt}
%\includegraphics[height=0.182\textwidth]{Rate_Region.eps}
%\caption{Rate region comparison. $p=5$ dB.}
%\label{Figure4}
%\vspace{-20pt}
%\end{figure}
\begin{figure}[!t]
    \centering
    \subfigbottomskip=0pt
	\subfigcapskip=-5pt
\setlength{\abovecaptionskip}{0pt}
    \subfigure[$\kappa=\mu=0.5$.]
    {
        \includegraphics[height=0.3\textwidth]{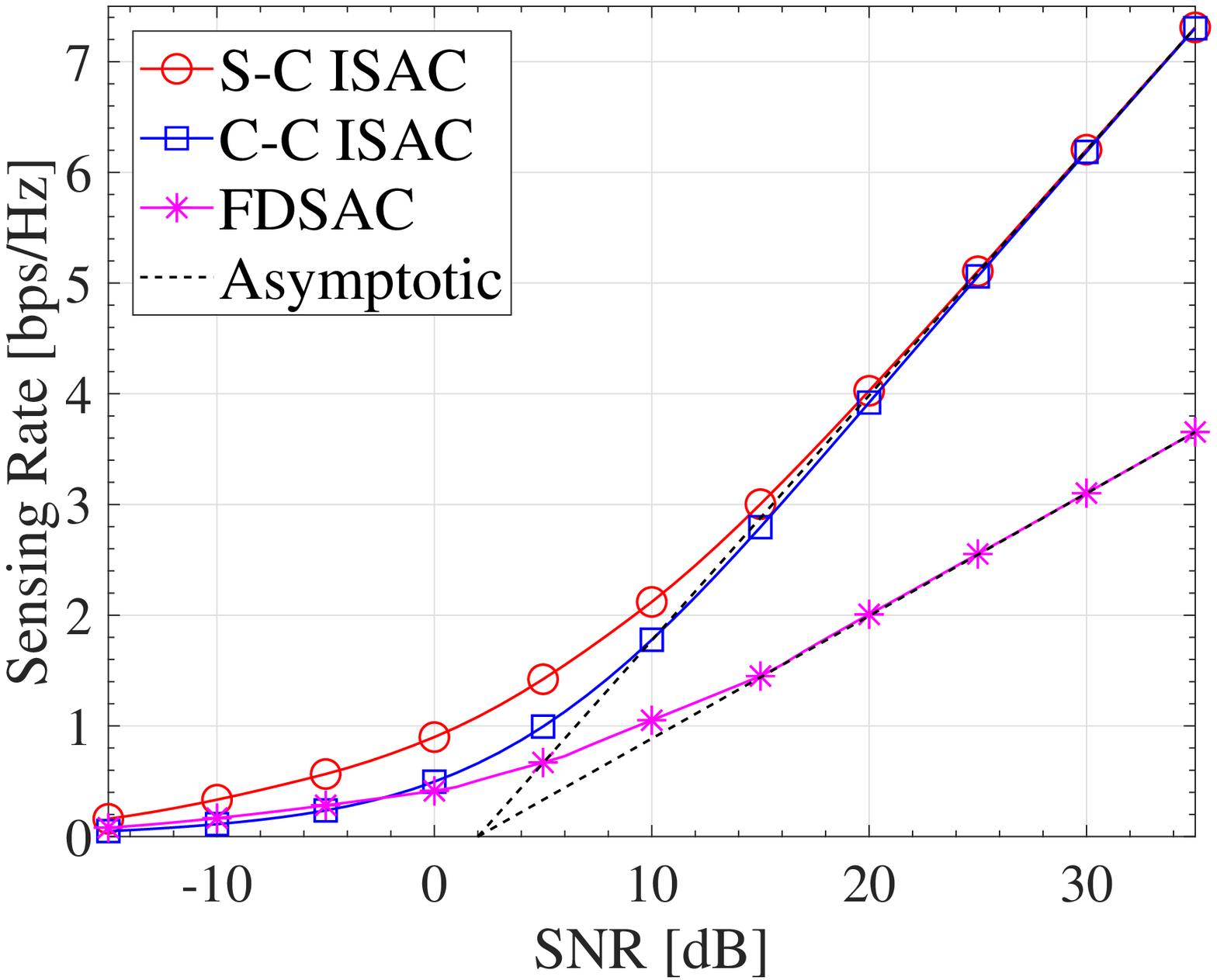}
	   \label{Figure3}	
    }
   \subfigure[$p=5$ dB.]
    {
        \includegraphics[height=0.3\textwidth]{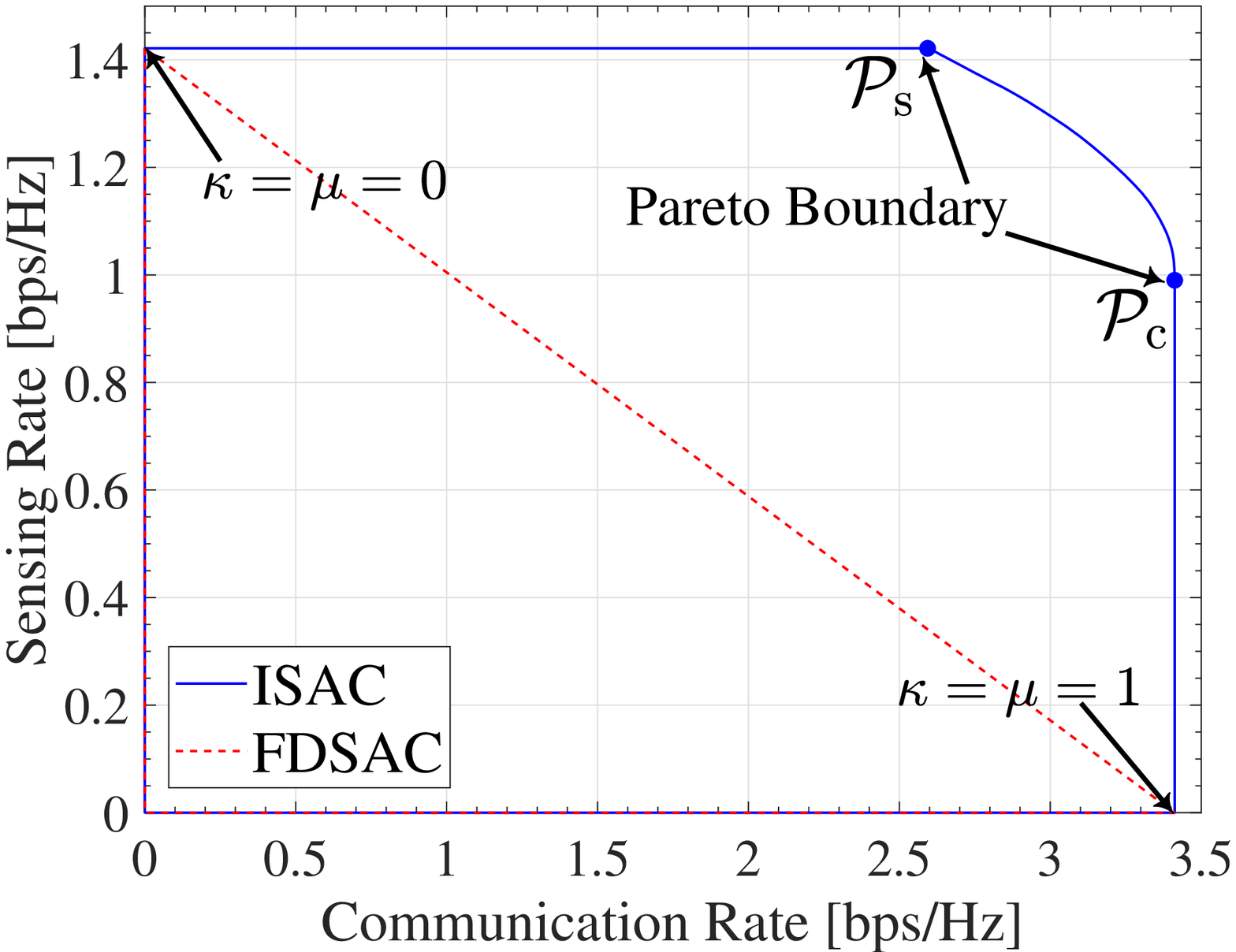}
	   \label{Figure4}	
    }
\caption{(a) Performance of sensing; (b) Rate region.}
    \vspace{-20pt}
\end{figure}

{\figurename} {\ref{Figure4}} compares the SR-CR regions achieved by ISAC (presented in \eqref{Rate_Regio_ISAC}) and FDSAC (presented in \eqref{Rate_Regio_FDSAC}). As shown, the rate region of FDSAC is plotted by changing the bandwidth allocation factor $\kappa$ and the power allocation factor $\mu$ from $0$ to $1$. As expected, a larger value of $\kappa$ or $\mu$ yields a higher CR. For ISAC, the point $\mathcal{P}_{\rm{s}}$ and the point $\mathcal{P}_{\rm{c}}$ are achieved by the S-C design and the C-C design, respectively. Moreover, the curve segment connecting $\mathcal{P}_{\rm{s}}$ and $\mathcal{P}_{\rm{c}}$ represents the Pareto boundary of ISAC's rate region, which is obtained by solving \eqref{Problem_CR_SR_Tradeoff} for $\alpha$ changing from 1 to 0. As {\figurename} {\ref{Figure4}} shows, the rate region achieved by FDSAC is completely included in that achieved by ISAC, which verifies the correctness of Theorem \ref{theorem_Rate_Region_Comparision}.
\vspace{-10pt}
\section{Conclusion}
In this letter, we have analyzed the S\&C performance of ISAC systems under three DFSAC precoding designs. The high-SNR slopes and diversity orders achieved by ISAC have been derived to highlight its superiority. Theoretical analyses have demonstrated that ISAC can provide more degrees of freedom and achieve a broader rate region than FDSAC.
\vspace{-10pt}
\begin{appendix}
%\subsection{Proof of Theorem \ref{Sensing_Rate_S_C_Theorem}}\label{Proof_Sensing_Rate_S_C_Theorem}
%It is noteworthy that maximizing ${\mathcal{R}}_{\rm{s}}$ is equivalent to maximizing the MI of a virtual MIMO channel $\dot{\mathbf{y}}={\mathbf{R}}^{1/2}\dot{\mathbf{x}}+\dot{\mathbf{n}}$ with ${\mathbbmss{E}}\{\dot{\mathbf{x}}{\dot{\mathbf{x}}}^{\mathsf{H}}\}={\mathbf{W}}{\mathbf{W}}^{\mathsf{H}}$ and $\dot{\mathbf{n}}\sim{\mathcal{CN}}\left({\mathbf{0}},L^{-1}{\mathbf{I}}_M\right)$. Thus, when ${\mathcal{R}}_{\rm{s}}$ is maximized, the eigenvectors of ${\mathbf{W}}{\mathbf{W}}^{\mathsf{H}}$ should equal the left eigenvectors of ${\mathbf{R}}^{1/2}$, with the eigenvalues chosen by the water-filling procedure, which yields $\mathcal{R}_{\rm{s}}^{\rm{s}}=\frac{N}{L}\sum_{m=1}^{M}\log_2\left(1+L\lambda_ms_m^{\star}\right)$ with $s_{m}^{\star}=\max\{0,1/{\nu}-{1}/(L\lambda_m)\}$ and $\sum_{m=1}^{M}s_{m}^{\star}=p$. When $p\rightarrow\infty$, we have $\nu\rightarrow0$ and $\sum_{m=1}^{M}s_m^{\star}=\frac{M}{\nu}-\sum_{m=1}^{M}\frac{1}{L\lambda_m}=p$. It follows that $\lim_{p\rightarrow\infty}s_m^{\star}\simeq\frac{p}{M}-\frac{1}{L\lambda_m}+\frac{1}{LM}\sum_{m=1}^{M}\frac{1}{\lambda_m}$, which together with the fact of $\lim_{x\rightarrow\infty}\log_2(1+x)\approx\log_2{x}$ yields $\lim_{p\rightarrow\infty}\mathcal{R}_{\rm{s}}^{\rm{s}}=\frac{NM}{L}\log_2{p}+\frac{N}{L}\sum_{m=1}^{M}\log_2\left(\frac{L\lambda_m}{M}\right)$.
%\vspace{-10pt}
\subsection{Proof of Theorem \ref{Ergodic_Communication_Rate_S_C_Theorem}}\label{Proof_Ergodic_Communication_Rate_S_C_Theorem}
Note that $s_{m}^{\star}$ is a constant and the probability density function (PDF) of $\rho_m=\lvert{\mathbf{v}}_{m}^{\mathsf{H}}{\mathbf{h}}_{m}\rvert^2$ is given by $f_{m}(x)=\frac{{\rm{e}}^{-x}}{K'!}x^{K'}$ \cite[Eq. (9.80)]{Heath2018}. On this basis, we can get \eqref{Ergodic_Communication_Rate_S_C_Basic} by \cite[Eq. (4.337.5)]{Ryzhik2007}. As stated before, $\lim_{p\rightarrow\infty}s_m^{\star}\simeq\frac{p}{M}\!-\!\frac{1}{L\lambda_m}\!+\!\sum_{m=1}^{M}\frac{1}{LM\lambda_m}$. This together with the fact of $\lim_{x\rightarrow\infty}\log_2(1\!+\!x)\approx\log_2{x}$ and \cite[Eq. (4.352.1)]{Ryzhik2007} yields \eqref{Ergodic_Communication_Rate_S_C_Asymptotic}.
\vspace{-10pt}
\subsection{Proof of Theorem \ref{Outage_Probability_UT_S_C_Theorem}}\label{Proof_Outage_Probability_UT_S_C_Theorem}
As stated before, $\lim_{p\rightarrow\infty}s_m^{\star}\simeq\frac{p}{M}-\frac{1}{L\lambda_m}+\frac{1}{LM}\sum_{m=1}^{M}\frac{1}{\lambda_m}$, which together with the fact of $\lim_{x\rightarrow\infty}\log_2(a+x)\approx\log_2{x}$ yields $\lim_{p\rightarrow\infty}{\overline{\mathcal{R}}}_{{\rm{c}}}^{\rm{s}}\simeq\log_2(\prod_{m=1}^{M}\frac{p}{M}\rho_m)$. Therefore, the OP satisfies $\lim_{p\rightarrow\infty}{{\mathcal{P}}}_{{\rm{c}}}^{\rm{s}}\simeq\Pr(\prod_{m=1}^{M}\frac{p}{M}\rho_m<2^{{\mathcal R}_0})$. On this basis, we can obtain $\lim_{p\rightarrow\infty}{{\mathcal{P}}}_{{\rm{c}}}^{\rm{s}}\simeq{\mathcal{O}}(p^{-M(K'+1)})$ by exploiting the approach in deriving \cite[Eq. (39)]{Chelli2014_TWC}.
\vspace{-10pt}
\subsection{Proof of Theorem \ref{Ergodic_Communication_Rate_C_C_Theorem}}\label{Proof_Ergodic_Communication_Rate_C_C_Theorem}
Clearly, as $p\rightarrow\infty$, we have $\upsilon\rightarrow0$ and thus $c_{m}^{\star}\simeq\frac{p}{M}-\frac{1}{\rho_m}+\frac{1}{M}\sum_{m=1}^{M}\frac{1}{\rho_m}$. It follows from the fact of $\lim_{x\rightarrow\infty}\log_2(a+x)\approx\log_2{x}$ that the ECR of UT $m$ satisfies $\lim_{p\rightarrow\infty}\mathbbmss{E}\{\log_2(1+\rho_mc_m^{\star})\}\simeq\log_2{\frac{p}{M}}+{\mathbbmss{E}}\{\log_2{\rho_m}\}$. Based on \cite[Eq. (4.352.1)]{Ryzhik2007}, the final results follow immediately.
\vspace{-10pt}
\subsection{Proof of Theorem \ref{theorem_Rate_Region_Comparision}}\label{Proof_theorem_Rate_Region_Comparision}
To proceed, we construct two auxiliary regions as follows:
{\setlength\abovedisplayskip{2pt}
\setlength\belowdisplayskip{2pt}
\begin{align}
&\mathcal{C}_{1}=\{\left({\mathcal{R}}^{\rm{s}},{\mathcal{R}}^{\rm{c}}\right)|{\mathcal{R}}^{\rm{s}}\!\in\!\left[0,\mathcal{R}_{{\rm{s}},1}^{\epsilon}\right],
{\mathcal{R}}^{\rm{c}}\!\in\![0,\mathbbmss{E}\{\overline{\mathcal{R}}_{{\rm{c}},1}^{\epsilon}\}],\epsilon\!\in\!\left[0,\!1\right]\},\nonumber\\
&\mathcal{C}_{2}=\{\left({\mathcal{R}}^{\rm{s}},{\mathcal{R}}^{\rm{c}}\right)|{\mathcal{R}}^{\rm{s}}\!\in\!\left[0,\mathcal{R}_{{\rm{s}},2}^{\epsilon}\right],
{\mathcal{R}}^{\rm{c}}\!\in\!\left[0,\mathcal{R}_{{\rm{c}},2}^{\epsilon}\right],\epsilon\!\in\!\left[0,\!1\right]\}.\nonumber
\end{align}
}Here,
{\setlength\abovedisplayskip{2pt}
\setlength\belowdisplayskip{2pt}
\begin{align}
\mathcal{R}_{{\rm{s}},1}^{\epsilon}\!=\!\frac{N}{L}\!\max\nolimits_{\sum_{m=1}^{M}\!k_m\leq(1-\epsilon)p}\!\sum\nolimits_{m=1}^{M}\!\log_2\left(1\!+\!L\lambda_mk_m\right) \end{align}
}and
{\setlength\abovedisplayskip{2pt}
\setlength\belowdisplayskip{2pt}
\begin{align}
\overline{\mathcal{R}}_{{\rm{c}},1}^{\epsilon}=\max\nolimits_{\sum_{m=1}^{M}u_m\leq\epsilon p}\sum\nolimits_{m=1}^{M}\!\log_2(1\!+\!u_m\rho_m).
\end{align}
}Let $\{k_m^{\epsilon}\}_{m=1}^{M}$ and $\{u_m^{\epsilon}\}_{m=1}^{M}$ denote the optimal solutions to $\{k_m\}_{m=1}^{M}$ and $\{u_m\}_{m=1}^{M}$ for a given $\epsilon$, respectively. Besides, $\mathcal{R}_{{\rm{s}},2}^{\epsilon}=\mathbbmss{E}\{\overline{\mathcal{R}}_{{\rm{s}},2}^{\epsilon}\}$ and $\mathcal{R}_{{\rm{c}},2}^{\epsilon}=\mathbbmss{E}\{\overline{\mathcal{R}}_{{\rm{c}},2}^{\epsilon}\}$, where $\overline{\mathcal{R}}_{{\rm{s}},2}^{\epsilon}$ and $\overline{\mathcal{R}}_{{\rm{c}},2}^{\epsilon}$ denote the SR and sum CR achieved by the precoding matrix $\mathbf{W}={\mathbf{U}}{\bm{\Delta}}_{\epsilon}^{1/2}$ with ${\bm{\Delta}}_{\epsilon}={\mathsf{diag}}\{k_1^{\epsilon}+u_1^{\epsilon},\ldots,k_M^{\epsilon}+u_M^{\epsilon}\}$, respectively. Clearly, we have $\mathcal{C}_{2}\subseteq{\mathcal{C}}_{\rm{i}}$ and ${\mathcal{C}}_{\rm{f}}\subseteq\mathcal{C}_{1}$. Furthermore, $\{(\mathcal{R}_{{\rm{s}},1}^{\epsilon},\mathbbmss{E}\{\overline{\mathcal{R}}_{{\rm{c}},1}^{\epsilon}\})|\epsilon\!\in\!\left[0,\!1\right]\}$ serves as the boundary of $\mathcal{C}_{1}$. Given $\epsilon_1\in[0,1]$, when $\overline{\mathcal{R}}_{{\rm{c}},1}^{\epsilon_1}\in[{{\mathcal{R}}}_{\rm{s}}^{\rm{c}},{\overline{\mathcal{R}}}_{\rm{c}}^{\rm{c}}]$, there exists an $\epsilon_2\in[0,1]$ with $\overline{\mathcal{R}}_{{\rm{c}},2}^{\epsilon_2}=\overline{\mathcal{R}}_{{\rm{c}},1}^{\epsilon_1}$. Using the monotonicity of function $\log_2(1+ax)$ ($a>0$) with respect to $x\geq0$, we obtain that $\epsilon_2\leq \epsilon_1$. By continuously using the monotonicity of $\log_2(1+ax)$, we can get $\overline{\mathcal{R}}_{{\rm{s}},2}^{\epsilon_2}\geq{\mathcal{R}}_{{\rm{s}},1}^{\epsilon_1}$. When $\overline{\mathcal{R}}_{{\rm{c}},1}^{\epsilon_1}\in[0,{{\mathcal{R}}}_{\rm{s}}^{\rm{c}}]$, we have $\overline{\mathcal{R}}_{{\rm{c}},1}^{\epsilon_1}\leq{{\mathcal{R}}}_{\rm{s}}^{\rm{c}}=\overline{\mathcal{R}}_{{\rm{c}},2}^{0}$ and $\overline{\mathcal{R}}_{{\rm{s}},1}^{\epsilon_1}\leq{{\mathcal{R}}}_{\rm{s}}^{\rm{s}}=\overline{\mathcal{R}}_{{\rm{s}},2}^{0}$. The above arguments imply that any rate-tuple on the boundary of ${\mathcal{C}}_1$ falls within ${\mathcal{C}}_2$ and thus ${\mathcal{C}}_{1}\subseteq\mathcal{C}_{2}$ holds. As stated before, $\mathcal{C}_{2}\subseteq{\mathcal{C}}_{\rm{i}}$ and ${\mathcal{C}}_{\rm{f}}\subseteq\mathcal{C}_{1}$. Taken together, we obtain $\mathcal{C}_{\rm{f}}\subseteq{\mathcal{C}}_{\rm{i}}$.
\vspace{-10pt}
\end{appendix}

\end{document}